\documentclass[aps,showpacs,superscriptaddress,prl,twocolumn]{revtex4}

\usepackage{amsmath}
\usepackage{graphicx}

\usepackage{amsmath}
\usepackage{amssymb}

\begin{document}

\title{Denaturation Transition of Stretched DNA}

\author{Andreas Hanke}
\author{Martha G. Ochoa}
\affiliation{Department of Physics and Astronomy, 
University of Texas at Brownsville, 80 Fort Brown, Brownsville, USA}
\author{Ralf Metzler}
\affiliation{Physics Department, Technical University of Munich,
D-85747 Garching, Germany}
\affiliation{Center for NanoScience, Ludwig Maximilians University, D-80539 Munich, 
Germany}

\begin{abstract}
We generalize the Poland-Scheraga model to consider DNA denaturation
in the presence of an external stretching force. We demonstrate the 
existence of a force-induced DNA denaturation transition and obtain
the temperature-force phase diagram. The transition 
is determined by the loop exponent $c$ for which we find the new value 
$c=4\nu-1/2$ such that the transition is second order with $c=1.85<2$ in 
$d=3$. We show that a finite stretching force $F$ destabilizes DNA, 
corresponding to a lower melting temperature $T(F)$, in agreement with 
single-molecule DNA stretching experiments.
\end{abstract}


\maketitle

Under physiological conditions the thermodynamically stable configuration
of DNA is the Watson-Crick double helix. The constituent
monomers of each helix, the nucleotides A, T, G, C, pair with those of
the complementary helix according to the key-lock principle, such that
only the base-pairs (bps) AT and GC can form \cite{kornberg}. Upon heating or
titration with acid or alkali of double-stranded DNA, regions of unbound
bps proliferate along the DNA until full separation of the two
DNA strands at the melting temperature $T_m$; depending on
the relative content of AT bps, $T_m$ ranges between some 60 to
110$^\circ$C \cite{delcourt}.
The classical Poland-Scheraga (PS) model views DNA as an alternating sequence
of intact double-helical and denatured, single-stranded domains (bubbles or
loops). Double-helical regions are dominated by the hydrogen bonding of bps
as well as base stacking, bubbles by the entropy gain on disruption of
bps \cite{ps}. The PS model is fundamental in biological
physics and has been progressively refined
to obtain a quantitative understanding of the DNA melting process
\cite{poland,wartell}.
DNA denaturation can also be induced mechanically, by longitudinal stretching
of single DNA molecules by optical or magnetic tweezers or atomic force 
microscopes \cite{williams,rief,bustamante,bensimon}.
At the transition, the plot of stretching force $F$ versus mean
DNA extension $L$ exhibits a plateau at 60-90\,pN
\cite{rouzina,prentiss}. Force-induced destabilization of DNA has become a
valuable
tool, e.g., to probe the interaction of proteins that specifically bind to
single-stranded DNA, at physiological melting temperatures $T_m(F)$ well 
below $T_m(0)$ of free DNA \cite{pant}.

\begin{figure}
\includegraphics[width=8.0cm]{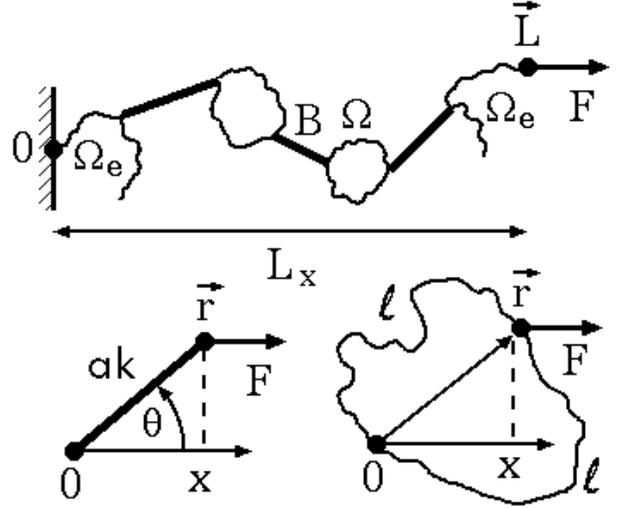}
\caption{Stretched DNA in the PS model with bound segments $B$ and
denatured loops $\Omega$. The DNA is attached between ${\cal O}$
and ${\bf L}$ and subject to the stretching force $F$ in 
$x$-direction. Perfect matching in heterogeneous DNA requires both 
arches of a loop to have equal length $\ell$.}
\label{model}
\end{figure}

\begin{figure}
\includegraphics[width=7.9cm]{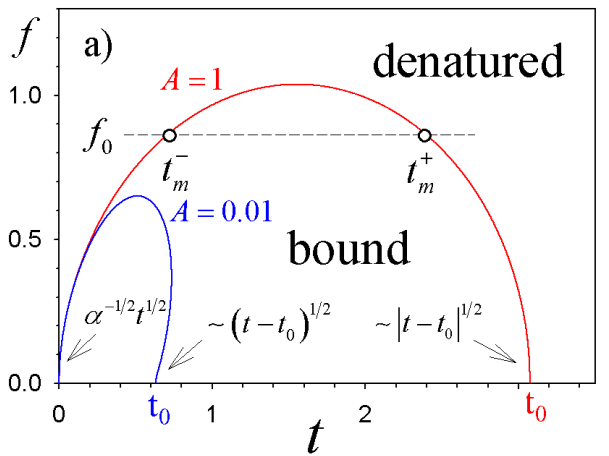}
\includegraphics[width=8.4cm]{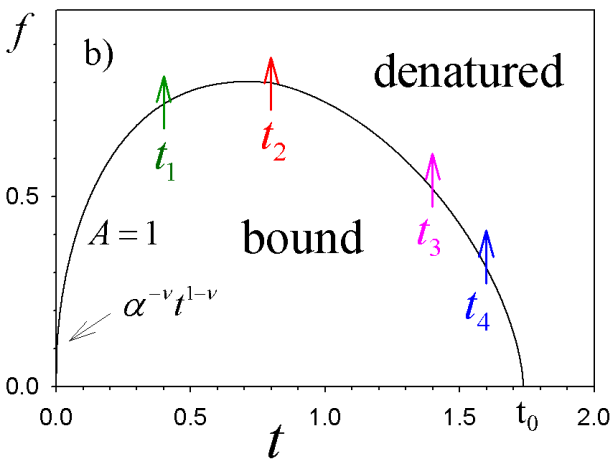}
\caption{Transition lines $f_m = F_m a / \varepsilon$ 
as function of $t = k_B T / \varepsilon$ for $\alpha = 1$, $s = 5$ 
for denatured loops modeled as (a) ideal random walks
and (b) self-avoiding walks (cf.~Fig.~\ref{fer}).}
\label{pd}
\end{figure}

\begin{figure}
\includegraphics[width=8.6cm]{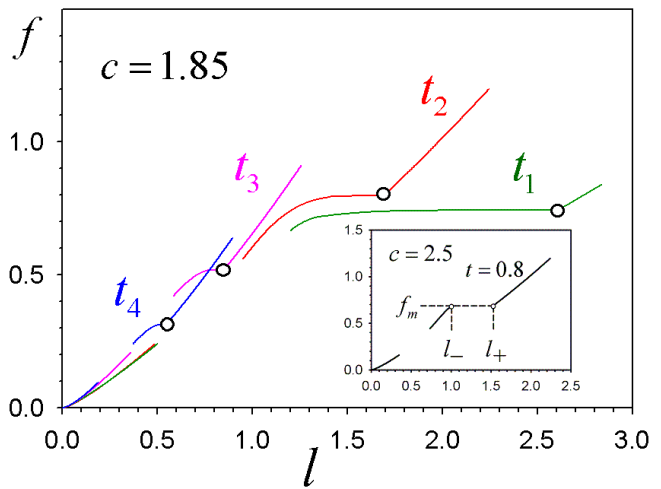}
\caption{Force-extension curves $f = F a / \varepsilon$ as function
of $l=\langle L_x\rangle/(a\langle N\rangle)$ at fixed
$t_1 < t_2 < t_3 < t_4$ (cf.~Fig.~\ref{pd}b). Open 
circles mark second-order transitions ($c = 1.85 < 2$).
Inset: $f(l)$ for $c = 2.5 > 2$ where the
transition is first order.}
\label{fer}
\end{figure}

In this Letter we consider the force-assisted denaturation transition of
double-stranded DNA in the framework of the PS model (Fig.~\ref{model}).
The thermodynamic state
of the DNA molecule now depends on both temperature $T$ and stretching force
$F$. We find a bounded region of bound states in the $(T,F)$ plane 
(Fig.~\ref{pd}). The shape of the transition line
implies that finite stretching forces $F$ indeed lower the melting
temperature $T_m(F)$, and the calculated force-extension relations $F(L)$
exhibit a plateau over a certain range of DNA extension $L$ (Fig.~\ref{fer}).
Both observations are in agreement with force-induced DNA melting experiments.

We treat the chain in the grand canonical ensemble in which the total number
$N$ of bps and the end-to-end vector ${\bf L}$ fluctuate. The partition
function in $d = 3$ becomes
\begin{equation}
\label{pf1}
\mathcal{Z}(z,F)=\sum_{N=1}^{\infty}\int d^3L
\mathcal{Z}_{\text{can}}(N,{\bf L})z^N\exp(\beta FL_x)
\end{equation}
with $\beta=1/(k_BT)$.
$\mathcal{Z}_{\text{can}}(N,{\bf L})$ is the canonical partition function
of a chain
of $N$ bps with fixed end-to-end vector ${\bf L}$ and $z$ is the fugacity.
We assume that the force $F$ acts in the positive $x$-direction, and $L_x$
is the $x$-component of ${\bf L}$ (Fig.~\ref{model}). 
If bound segments and bubbles are independent, $\mathcal{Z}$ factorizes:
\begin{equation}
\label{pf2}
\mathcal{Z}(z,F)=\Omega_e+\Omega_e\left\{\sum\limits_{n=0}^{\infty}
\left[B\Omega\right]^n\right\}B\Omega_e\,\,,
\end{equation}
the last term equaling $\Omega_e^2B/(1-B\Omega)$. The alternating sequence
of bound segments and bubbles with weights $B$ and $\Omega$ in Eq.~(\ref{pf2})
is complemented by the weight $\Omega_e$ of an open end unit at both ends of
the chain. Note that only
one strand of the end unit is bound to the, say, magnetic bead, while the 
other strand is
moving freely. 

We model a \emph{bound segment\/} with $k=1,2,\ldots$ bps as a rigid
rod of length $ak$ where $a=0.34$\,nm is the length of a bound bp
in B-DNA \cite{rouzina}. For simplicity we assume that the binding
energy $E_0<0$ per bp is the same for all bps. The statistical
weight of a segment with fixed number $k$ and fixed orientation is
then $\omega^k$ with $\omega=\exp(\beta \varepsilon)$ and $\varepsilon
=-E_0>0$. Assuming that $k$ fluctuates with fixed fugacity
$z$, and rotates around one end while subject to the force $F$
(Fig.~\ref{model}), the statistical weight of the segment for
fixed $z$ and $F$ becomes
\begin{subequations} \label{B}
\begin{eqnarray}
\label{B1}
B(z,\omega,F)&=&\sum\limits_{k=1}^{\infty} \frac{(\omega z)^k}{4 \pi}
\int\limits_{\Omega} d\Omega \, \exp(\beta F x)\\
&=&\frac{1}{2 y}\ln\left(\frac{1-\omega ze^{-y}}{1-\omega ze^y}\right),
\,\,y\equiv\beta Fa.
\label{B2}
\end{eqnarray}
\end{subequations}
Integration in Eq.~(\ref{B1}) is over the unit sphere with area $4\pi$,
and $x=ak\cos\theta$ where $\theta$ is the polar angle between segment and
$x$-axis. At $F=0$, $B(z,\omega,0)=\omega z/(1-\omega z)$ as found previously
for the denaturation transition
of free DNA \cite{kafri}. Note that $B$ is only well-defined for $\omega
ze^y<1$; in what follows we assume $z<e^{-y}/\omega$.

\emph{Denatured loops\/} are considered as closed random walks with $2\ell$
monomers, corresponding to $\ell$ broken bps. This loop starts at
${\cal O}$ and visits the point ${\bf r}$ after $\ell$ monomers
(Fig.~\ref{model}). The number of configurations of a loop is
\begin{equation}
\label{Olr}
\Omega(\ell,{\bf r})=C_0(2\ell)p_{\ell}({\bf r})
\end{equation}
under this constraint,
where $C_0(2 \ell)$ counts the configurations of a loop of length $2\ell$
starting at ${\cal O}$ and $p_{\ell}({\bf r})$ is the probability
that the loop visits ${\bf r}$ after $\ell$ monomers. For an ideal 
random walk in $d=3$, $C_0(2\ell)\sim\mu^{2\ell}\ell^{-3/2}$
($\mu$ is the connectivity constant) and $p_{\ell}({\bf r})\sim{\cal R}^{-3}
\exp[-\lambda(r/{\cal R})^2]$ where $\lambda > 0$, $r=|{\bf r}|$,
and ${\cal R}=b\ell^{1/2}$ is the scaling length of the walk.
The amplitude $b$ is proportional to the persistence
length of the walk. Thus, $\Omega(\ell,{\bf r})\sim
s^{\ell}\ell^{-3}\exp[-\lambda(r/{\cal R})^2]$ where $s=\mu^2$. We assume
that ${\bf r}$ moves freely and is subject to the force $F$ in the positive
$x$-direction. The weight of an ideal random loop for fixed
$\ell$ and $F$ is given by the Gaussian integral
\begin{equation}
\label{O1}
\Omega(\ell,F)=\int d^3r\,\Omega(\ell,{\bf r})\,e^{\beta Fx}
=As^{\ell}\ell^{-c}\exp({\alpha y^2\ell})
\end{equation}
where $A$ is an amplitude, $c = 3/2$, and $\alpha=b^2/(4\lambda a^2)$.
Finally, we sum $\Omega(\ell,F)$ over $\ell$ with weight $z^{\ell}$
to obtain the statistical weight for an ideal random loop
\begin{equation}
\label{OzF1}
\Omega(z,F)=A\sum\limits_{\ell=1}^{\infty}u^{\ell}\ell^{-c}
=A\mbox{Li}_c(u),\,\,u=sz\exp(\alpha y^2).
\end{equation}
$\mbox{Li}_c(u)=\sum_{\ell=1}^{
\infty}u^{\ell}\ell^{-c}$ is the polylog function \cite{wolfram},
converging for $|u|<1$ for any $c$. For $u=1$ three cases exist:
(i) $c\leq1$: $\mbox{Li}_c(1)$ diverges; (ii) $1<c\leq2$: $\mbox{Li}_c(1)$
converges but $\left.\mbox{Li}_c'(u)\right|_{u=1}$ diverges; (iii) $c>2$: Both
$\mbox{Li}_c(1)$ and $\left.\mbox{Li}_c'(u)\right|_{u=1}$ converge. The limit
$u=1$ corresponds to the value $z_m(F)=\exp(-\alpha y^2)/s$ of the fugacity;
thus, $\Omega(z,F)$ is only well-defined for $z\le z_m(F)$ and diverges for
$z>z_m(F)$. The statistical weight $\Omega_e$ of an end unit modeled as ideal
random walk may be derived in a similar way and one obtains
$\Omega_e(z,F)=A_{e}\mbox{Li}_{0}(u)$.

For free DNA it was found that the nature of the denaturation transition is
determined by the analytic behavior of $\mbox{Li}_c(u)$ at $u=1$:
for $c\leq1$ there is no phase transition in the thermodynamic sense; for $1<
c\leq2$ the transition is second order, and for $c>2$ it is first order
\cite{ps,kafri}. One finds $c=3/2<2$ if the loops are ideal random walks.
Self-avoiding interactions within a loop modify this value to $c=3\nu=
1.76$ with $\nu=0.588$ in $d=3$ \cite{Fisher1966}. In both cases the
transition is second order. Self-avoiding interactions between denatured loops
and the rest of the chain were found to produce
$c=2.12>2$, driving the transition to first order \cite{kafri,carlon}.
These results suggest that the inclusion of self-avoiding interactions
generally shifts the loop exponent $c$ to larger values, possibly effecting
a change of the transition from second to first order.

To see how $c$ changes when self-avoiding interactions within a loop are
included for the case $F>0$, we obtain the weights $\Omega$ and $\Omega_e$
for a self-avoiding walk for $\ell \to \infty$. Then, Eq.~(\ref{Olr})
holds with $C_0(2\ell)\sim\mu^{2\ell}\ell^{-d\nu}$ being the number of
self-avoiding loops with $2\ell$ monomers. The probability density 
$p_{\ell}({\bf r})$ scales as $p_{\ell}({\bf r})={\cal R}^{-d}g(r/{\cal R})$
where ${\cal R}=b\ell^{\nu}$ is the scaling length of a self-avoiding walk
and $g(x)$ a scaling function. The function $g(x)$ is not
known for a self-avoiding loop. In what follows we assume
$g(x)\sim x^{\phi}\exp[-\lambda x^{\delta}]$ for $x\to\infty$ 
where $\lambda > 0$,
$\phi$ is an exponent, and $\delta=1/(1-\nu)$ is determined by
an argument by Fisher \cite{Fisher1966}. This
form of $g(x)$ is consistent with $p_{\ell}({\bf r})$ for a Gaussian loop
($\nu=1/2$, $\phi=0$) obtained above. For the related {\em linear\/}
self-avoiding walk starting at ${\cal O}$ and ending at ${\bf r}$ after
$\ell$ monomers, the above form of $g(x)$ also holds and ${\phi}$ can be
expressed in terms of known exponents \cite{Fisher1966,KM1971}. 
For the present case of a
self-avoiding loop $\delta=1/(1-\nu)$ still holds but ${\phi}$
is unknown. However, we will see that $\phi$ drops out from the result for
$\Omega(z,F)$ in the limit $\ell\to\infty$ at $F>0$.
The integral in Eq.~(\ref{O1}) is no longer
Gaussian, but can be evaluated using the steepest descent method at
$\kappa=\beta Fb\ell^{\nu}\to\infty$. It turns out that
in this limit the integral is dominated by values 
$r/\ell^{\nu}\to\infty$. With the above behavior of
$g(x)$ at $x\to\infty$ we find for a self-avoiding loop 
[cf.~Eq.~(\ref{O1})]
\begin{equation}
\label{O4}
\Omega(\ell,F)=As^{\ell}\ell^{-c}y^{1/(2\nu)-1}
\exp\left(\alpha y^{1/\nu}\ell\right)
\end{equation}
for $\kappa \to \infty$
with the new loop exponent in $d = 3$,
\begin{equation}
\label{c}
c=4\nu-1/2=1.85 \, \, .
\end{equation}
Thus, with self-avoiding interactions within a denatured loop
and $F>0$ the transition remains
second order, but moves closer to first order
compared to free DNA (with $c=3\nu=1.76$ obtained within the same approach).
The amplitude $A$ in Eq.~(\ref{O4}) is proportional to the cooperativity
parameter $\sigma_0\ll 1$ quantifying the initiation of a loop in a previously
intact double strand in the PS model \cite{poland,wartell}, such that also
$A\ll1$. Moreover, $\alpha = 0.6\ldots1.7 \approx 1$ using 
$\alpha=b^2/(4\lambda a^2)$ obtained for an ideal random walk where 
$b^2 / \lambda = 2 L_p x_{ss} / 3$; here, $x_{ss} = 0.6$\,nm is the length 
of a base in single-stranded DNA \cite{rouzina} and values for the persistence 
length $L_p$ for single-stranded DNA were found to range between
0.7\,nm \cite{bustamante} and 2\,nm \cite{murphy}.

Finally, we sum $\Omega(\ell,F)$ over $\ell$ with weight $z^{\ell}$
to obtain the statistical weight for a self-avoiding loop
[cf.~Eq.~(\ref{OzF1})],
\begin{equation}
\label{Omega}
\Omega(z,F)=Ay^{-\theta}\mbox{Li}_c\left[sz\exp(\alpha y^{1/\nu})\right],
\end{equation}
where $\theta=1-1/(2\nu)=0.15$ in $d=3$. The critical fugacity
is now given by $z_m(F)=\exp(-\alpha y^{1/\nu})/ s$. The weight $\Omega_e(
z,F)$ for an end unit obtains similarly, the result being Eq.~(\ref{Omega})
with $c$ replaced by $\zeta =3/2+\nu-2\gamma=-0.232$, using $\gamma=1.16$.

{\it Phase diagram.} We now obtain the transition line between bound and
denatured states in the $(T,F)$-plane in the thermodynamic limit $N\to\infty$.
For given fugacity $z$ the average number of bps (open and closed) becomes
\begin{equation}
\label{N}
\langle N\rangle=\partial\ln \mathcal{Z}(z,\omega,F)/\partial\ln z,
\end{equation}
where we explicitly include the argument $\omega$ from Eq.~(\ref{B}) in the
partition function (\ref{pf2}). If $N$ is set one has to
choose a fugacity $z$ such that $N=\langle N \rangle$; in this case $z$
becomes a function of $\omega$, $F$, and $N$. We denote the value of $z$
in the limit $N\to\infty$ by $z^*(\omega, F)\equiv\lim\limits_{N\to
\infty}z(\omega,F,N)$. Similar to the case $F=0$ \cite{kafri}, $z^*(\omega,F)$
is the lowest value of $z$ for which expression (\ref{N}) diverges.
In the bound state the divergence turns out to occur when the
denominator in $\Omega_e^2B/(1-B\Omega)$ vanishes 
[see text below Eq.~(\ref{pf2})], implying $z^*(\omega,F)$ to
satisfy
\begin{equation}
\label{bound}
B[z^*,\omega,F]\Omega[z^*,F]=1\,,\quad\mbox{bound state} \, .
\end{equation}
Conversely, in the denatured state the divergence occurs
because $\partial_z \Omega_e(z,F)$ diverges, which implies
\begin{equation}
\label{den}
z^*(\omega,F)=z_m(F)\,,\quad\mbox{denatured state}\, ,
\end{equation}
where $z_m(F)$ is the critical fugacity obtained above
(which is independent of $\omega$). Thus, starting in
a bound state in the $(T,F)$-plane and approaching the transition line by
varying $T$ and $F$, the value $z^*(\omega,F)$ is determined by
Eq.~(\ref{bound}) and increases until it reaches the value $z_m(F)$ from
Eq.~(\ref{den}). At this point the denaturation transition occurs. In the
denatured state $z^*(\omega,F)$ is given by Eq.~(\ref{den}).
Right at the transition both Eqs.~(\ref{bound}) and (\ref{den})
hold simultaneously. Using $\Omega[z_m(F),F]=Ay^{-\theta}\mbox{Li}_c(1)$ 
by definition of $z_m(F)$ this implies 
$A(\beta Fa)^{-\theta}\mbox{Li}_c(1)=1/B[z_m(F),\omega,F]$, 
relating $F$ and $\omega$, or, equivalently, the reduced force
$f=Fa/\varepsilon$ and temperature $t=k_BT/\varepsilon$, for the transition
line in the $(t,f)$-plane.

The shape of the transition line $f_m(t)$ depends on $A$, $\alpha$,
and $s$. Fig.~\ref{pd}a shows $f_m(t)$ for $A=1$, $\alpha=1$, and
$s=5$ for the case that denatured loops are ideal random walks
($\theta=0$, $\nu=1/2$). The transition line for the more realistic
value $A\ll1$ is also shown (here $A=0.01$). 
The line $f_m(t)$ separates a finite region of bound states from an
infinite region of denatured states. The point $(t_0,f=0)$ with $t_0=t_m(f
=0)$ corresponds to the traditional melting transition for free DNA ($F=0$).
The line $f_m(t)$ for $A=1$ contains a region in which
$f_m(t)$ {\em decreases\/} with $t$, such that increased stretching forces
$f$ lower the melting temperature $t_m(f)$, corresponding to force-induced
destabilization of DNA \cite{rouzina}. Interestingly, for $A=0.01$,
application of a small stretching force $f$ first
{\em increases} $t_m$ \cite{lee,williams2}. 
Moreover, $f_m(t)$ vanishes for both $t\to t_0$ 
(as $|t-t_0|^{1/2}$) and $t\to 0$ (as $\alpha^{-1/2}t^{1/2}$). 
This means that for given $0<f_0<f_{\text{
max}}$, where $f_{\text{max}}$ is the maximum of $f_m(t)$, 
the chain does not only denature at a large $t_m^+(f_0)$ but
also at a small $t_m^-(f_0)$, as indicated in \cite{mao}.
This behavior can be traced back to a balance
of the terms $(\beta F a)^2$ and $\beta F a$ in $z_m(F)=\exp(-\alpha y^2)/s$
and Eq.~(\ref{B2}), respectively \cite{interplay}.
For $(\beta F a)^2\ll\beta F a$, i.e., $k_BT
\gg Fa$, the melting transition at $t_m^+(f_0)$ is mainly driven by the
entropy gain on creation of fluctuating loops, similar as for 
free DNA. For $k_BT\ll Fa$ the transition at $t_m^-(f_0)$ is due
to the fact that $B[z_m(F),\omega,F]$ {\em decreases} with 
$y = \beta F a = f/t$ in the denatured state, due to the 
rapid decay of $z_m(F)$ [cf.~Eq.~(\ref{B2})] \cite{note}.
Fig.~\ref{pd}b shows the line $f_m(t)$ for
self-avoiding loops with $A=1$ and $c = 1.85$. 
Note that Eq.~(\ref{Omega})
reduces to the known result for a free self-avoiding loop ($y=0$)
only if $\theta=0.15$ is replaced by $\theta=0$; this is not
a contradiction since Eq.~(\ref{Omega}) is based on
the assumption that $\kappa=\beta bF\ell^{\nu}$ is large.
To include in Fig.~\ref{pd}b the behavior of $f_m(t)$
for $f=Fa/\varepsilon\to0$ we use Eq.~(\ref{Omega}) with
$\theta=0.15$ for $y>1$ and $\theta=0$ for $y\le1$.

{\it Force-extension relations.} In thermal denaturation of DNA one measures
the fraction $\Theta$ of bound bps as function of $T$. From the partition
function (\ref{pf1}) the average number $\langle M\rangle$ of bound bps is
$\langle M\rangle=\partial\ln\mathcal{Z}/\partial\ln\omega$ and
$\Theta=\langle M\rangle/\langle N\rangle$ with $\langle N \rangle$ 
from Eq.~(\ref{N}). 
Conversely, stretching experiments on DNA reveal its response to an applied
mechanical stress. The mean of the component of the DNA extension
along $F$ is
$\langle L_x\rangle=\beta^{-1}\partial\ln \mathcal{Z}/\partial F$.
The average extension per bp in units of the bp-bp distance $a$ is $l=
\langle L_x\rangle/(a\langle N\rangle)$. For comparison with experiments and
simulations we calculate $l$ in the thermodynamic limit $\langle N\rangle
\to\infty$. Consider the Gibbs-Duhem relation for the thermodynamic potential
$\ln \mathcal{Z}(z,\omega,F)$ \cite{volume}: $Nd\ln z+Md\ln\omega+
\beta L_xdF=0$. If $N$ is fixed one obtains $d\ln z+
\Theta d\ln\omega+ldy=0$ where $z$ is a function of
$\omega$, $F$, and $N$. For $N\to\infty$,
$d\ln z^*+\Theta d\ln\omega+ldy=0$,
$z^*(\omega, F)$ being the fugacity for $\langle N\rangle\to
\infty$ as discussed above; $\Theta(\omega,F)$ and $l(\omega,F)$ are 
the bound bp fraction and reduced DNA extension in the same limit.
For constant $y=\beta F a$ (or $y=0$) one finds
$\Theta=-\partial\ln z^*(\omega,F)/\partial\ln\omega$
(so that $\Theta=0$ in the denatured state due to Eq.~(\ref{den})
as expected). For constant $\omega$, corresponding to constant 
$t = k_B T / \varepsilon$, we find
$l(\omega,F)=-(a\beta)^{-1}\partial\ln z^*(\omega,F)/\partial F$.
Based on this result Fig.~\ref{fer} shows force-extension relations
$f(l)$ at fixed $t_1<t_2<t_3<t_4$
for the case that denatured loops are self-avoiding random walks
(cf.~Fig.~\ref{pd}b). The two sets of curves correspond to 
expansions of $f(l)$ for small $f$ and close to the transition,
respectively. The curves $f(l)$ display flattened regions close to 
the transition, in qualitative agreement with experimental force-extension 
relations for DNA. These regions become less pronounced as $t$ increases 
and vanish for $t\to t_0=t_m(F=0)$. A force-extension
relation for the case $c>2$, for which the transition is first order, is also
shown (here $c=2.5$). In this case $l(f)$ jumps discontinuously
from a value $l_{-}$ to a
larger value $l_{+}$ at the transition.

We have shown that a longitudinal stretching force $F$ results in 
a reduced denaturation temperature $T_m(F)$,
corresponding to force-induced destabilization of DNA.
For the loop exponent in the presence of a finite $F>0$ we found 
$c=4\nu-1/2=1.85$, so that the denaturation transition remains 
second order,
but with an increased exponent. It would be interesting to study how the
value of $c$ is modified when self-avoiding interactions between
a loop and the rest of the chain are included \cite{kafri,carlon}.

This work was supported by the NIH through SCORE grant 
GM068855-03S1 and by the AFOSR through grant FA9550-05-1-0472
(AH and MGO).

\vspace*{-3mm}



\begin{thebibliography}{99}

\bibitem{kornberg} A. Kornberg and T.~A. Baker, \emph{DNA Replication}
(W.~H.~Freeman, New York, 1992).

\bibitem{delcourt} S.~G. Delcourt and R.~D. Blake, J. Biol. Chem. \textbf{266},
15160 (1991).

\bibitem{ps} D. Poland and H.~A. Scheraga, J. Chem. Phys. \textbf{45},
1456 (1966).

\bibitem{poland} D. Poland and H.~A. Scheraga,
{\em Theory of Helix-Coil Transitions in Biopolymers}
(Academic, New York, 1970).

\bibitem{wartell} R.~M. Wartell and A.~S. Benight,
Phys. Rep. {\bf 126}, 67 (1985).

\bibitem{bustamante} S.~B. Smith, Y.~J. Cui, and C. Bustamante, Science
\textbf{271}, 795 (1996).

\bibitem{williams} J.~R. Wenner, M.~C. Williams, I. Rouzina, and V.~A.
Bloomfield, Biophys. J. \textbf{82}, 3160 (2002).

\bibitem{rief} H. Clausen-Schumann, M. Rief, C. Tolksdorf, and H.~E. Gaub,
Biophys. J. \textbf{78}, 1997 (2000).

\bibitem{bensimon} T.~R. Strick {\em et al.},
Rep. Prog. Phys. \textbf{66}, 1 (2003).

\bibitem{rouzina} I. Rouzina and V.~A. Bloomfield, Biophys. J. \textbf{80},
882 (2001).

\bibitem{prentiss} C. Limouse {\em et al.}, E-print arXiv:0704.3753v1.

\bibitem{pant} K. Pant, R.~L. Karpel, and M.~C. Williams, J. Mol. Biol.
\textbf{327}, 571 (2003);
I.~M. Sokolov, R. Metzler, K. Pant, and M.~C. Williams,
Biophys. J. \textbf{89}, 895 (2005).

\bibitem{kafri} Y. Kafri, D. Mukamel, and L. Peliti, Phys. Rev. Lett.
{\bf 85}, 4988 (2000); Eur. Phys. J. B {\bf 27}, 132 (2002).

\bibitem{wolfram} See http://functions.wolfram.com.

\bibitem{Fisher1966} M.~E. Fisher, J. Chem. Phys. {\bf 44}, 616 (1966).

\bibitem{carlon} E. Carlon, E. Orlandini, and A.~L. Stella,
Phys. Rev. Lett. {\bf 88}, 198101 (2002).

\bibitem{KM1971} D.~S. McKenzie and M.~A. Moore, J. Phys. A {\bf 4},
L82 (1971).

\bibitem{murphy} M.~C. Murphy {\em et al.}, Biophys. J. \textbf{86}, 2530 (2004).

\bibitem{lee} J.-B. Lee {\em et al.}, Nature (London) {\bf 439}, 621 (2006).

\bibitem{williams2} M.~C. Williams, J.~R. Wenner, I. Rouzina, 
and V.~A. Bloomfield, Biophys. J. \textbf{80}, 1932 (2001).

\bibitem{mao} H. Mao {\em et al.}, Biophys. J. \textbf{89}, 1308 (2005).

\bibitem{interplay} One may study the interplay between
$(\beta F a)^2$ and $\beta F a$ explicitly in a simplified model in which 
bound segments are always aligned along the $x$-direction. This produces 
a quadratic equation for $f_m$ with $t$ as a parameter, and $f_m(t)$ is 
a combination of both branches of the solution.

\bibitem{note} This relies on the assumption that
$p_{\ell}({\bf r})$ is {\em Gaussian} for ideal random 
loops and as described above Eq.~(\ref{O4}) for self-avoiding loops.
For very large $\beta F$ denatured loops are stretched out and
aligned along $F$ so that the partition function is dominated 
by parameter values for which $p_{\ell}({\bf r})$ deviates from this 
form. A suitable $p_{\ell}({\bf r})$ should be used to obtain the phase
diagram in this regime.

\bibitem{volume} As such the potential $\ln Z(z,\omega,F)$ depends only on
intensive parameters and would vanish due to the Euler equation.
This is avoided by formally including the system volume ${\cal V}$
in the potential. Here ${\cal V}=\infty$ is understood.

\end{thebibliography}
\end{document}